# Optimal power allocation for downstream xDSL with per-modem total power constraints : Broadcast Channel Optimal Spectrum Balancing (BC-OSB)

Vincent Le Nir, Marc Moonen, Jan Verlinden, Mamoun Guenach

*Abstract—*

*Recently, the duality between Multiple Input Multiple Output (MIMO) Multiple Access Channels (MAC) and MIMO Broadcast Channels (BC) has been established under a total power constraint. The same set of rates for MAC can be achieved in BC exploiting the MAC-BC duality formulas while preserving the total power constraint. In this paper, we describe the BC optimal power allocation applying this duality in a downstream x-Digital Subscriber Lines (xDSL) context under a total power constraint for all modems over all tones. Then, a new algorithm called BC-Optimal Spectrum Balancing (BC-OSB) is devised for a more realistic power allocation under per-modem total power constraints. The capacity region of the primal BC problem under per-modem total power constraints is found by the dual optimization problem for the BC under per-modem total power constraints which can be rewritten as a dual optimization problem in the MAC by means of a precoder matrix based on the Lagrange multipliers. We show that the duality gap between the two problems is zero. The multi-user power allocation problem has been solved for interference channels and MAC using the OSB algorithm. In this paper we solve the problem of multi-user power allocation for the BC case using the OSB algorithm as well and we derive a computational efficient algorithm that will be referred to as BC-OSB. Simulation results are provided for two VDSL2 scenarios: the first one with Differential-Mode (DM) transmission only and the second one with both DM and Phantom-Mode (PM) transmissions.*

## I. INTRODUCTION

In 1996, Foschini and Telatar have shown that the capacity of Multiple Input Multiple Output (MIMO) systems increases linearly with the minimum number of transmitters and receivers [2], [3]. In multi-carrier systems with Channel State Information (CSI) at the transmit and receive sides, the optimal capacity is reached using standard waterfilling [3]. In this case, the precoding and equalization are given by unitary matrices calculated from the Singular Value Decomposition (SVD) of the MIMO channel for each subcarrier. In [4], this work has been extended to the multi-user case, where the optimal transmit vector covariance matrices for the MIMO MAC are found by a procedure called iterative waterfilling. This

V. Le Nir and M. Moonen are with the SISTA/ESAT laboratory, Katholieke Unversiteit Leuven, Leuven, Belgium. E-mail: vincent.lenir@esat.kuleuven.be marc.moonen@esat.kuleuven.be

J. Verlinden and M. Guenach are respectively with Alcatel-Lucent and Alcatel-Lucent Bell Labs, Antwerpen, Belgium.

procedure searches the optimal power allocation iteratively over the users using the standard waterfilling formulas over frequencies.

However, finding the optimal transmit vector covariance matrices for the MIMO Broadcast Channels (BC) has been an open problem for a while. Indeed, contrary to the Single Input Single Output (SISO) BC where the channel is degraded (i.e each signal is a linear combination of the other signals and an additional noise), the MIMO BC is non-degraded which makes the capacity regions much more difficult to characterize. In fact, the SISO BC capacity region can be achieved by superposition coding at the transmit side and successive decoding at the receive side where the ordering between users is determined by the noise variance [5]. Recently, a new technique called Dirty Paper Coding (DPC) has been introduced [6], [7]. It can be shown that DPC achieve the capacity region for SISO and MIMO BC. Moreover, an important result has been established in the form of a duality theory between the MAC and BC, where the MIMO BC capacity regions can be characterized by their dual MIMO MAC capacity regions [8]. With this duality between the MAC and BC, it is found that the same set of rates can be obtained in both domains under the same total power constraint.

However, in a multi-user scenario, and especially in xDSL, a constraint on the total power used by each individual modem is more realistic than a total power constraint for all modems together. In a recent paper, per-modem total power constraints have been applied to MIMO MAC-BC duality theory in a wireless context [9]. It has been shown that MIMO MAC-BC duality still holds if an unknown covariance matrix is included in the MIMO MAC optimization function. The capacity regions are found by means of a maximization on the input covariance matrices and a minimization on the unknown covariance matrix, requiring complex algorithms in order to find the optimal solution when applied to the multi-tone transmission.

In this paper we devise a new algorithm called BC-OSB requiring less complexity compared to [9] for optimal power allocation under per-modem total power constraints for the multi-tone case (xDSL context). Our paper extends the work of [8] in practical scenarios where we have per-modem total power constraints for the BC. Indeed, it is known from [8] that per-modem total power constraints are not preserved by the MAC-BC transformations. If a single total power budget is used as in [8], then the power budget on each modem could be exceeded. Therefore we design a precoder for the BC based on

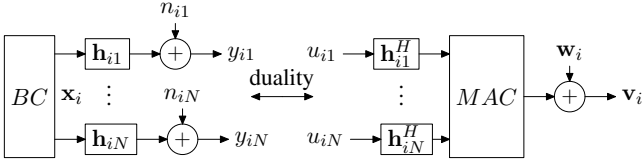

Fig. 1. BC (left side) and its dual MAC (right side) for tone $i$

Lagrange multipliers in order to transform these per-modem total power constraints into a virtual total power constraint and finally use MAC-BC transformations. This precoder rescales the channel matrix in order to meet per-modem total power constraints in the BC. The capacity region of the primal BC problem under per-modem total power constraints is found by the dual optimization problem for the BC under per-modem total power constraints which can be rewritten as a dual optimization problem in the MAC by means of a precoder matrix based on the Lagrange multipliers. We show that the duality gap between the two problems is zero.

The paper is organized as follows. In section II, we first introduce the BC optimal power allocation applying this duality in a downstream x-Digital Subscriber Lines (xDSL) context under a total power constraint for all modems over all tones. Then, a new algorithm called BC-OSB is devised in section III for a more realistic power allocation under per-modem total power constraints. Simulation results are given in section IV for a VDSL2 scenario with Differential-Mode (DM) transmission and a VDSL2 scenario with both DM and Phantom-Mode (PM) transmission.

## II. BC POWER ALLOCATION UNDER A TOTAL POWER CONSTRAINT

In this section, we describe a procedure for computing optimal transmit vector covariance matrices in the BC using MAC-BC duality under a total power constraint, that is under a single total power constraint for all tones and all modems in the MIMO system. This provides a reference for the derivations in section III.

We assume that all modems are synchronized and use Discrete Multi-Tone (DMT) modulation with a cyclic prefix longer than the maximum delay spread of the channel. We consider a MIMO Broadcast Channel (BC) serving N users in a xDSL downstream scenario as shown on the left side of Fig. 1. The transmission on one tone can then be modelled as:

$$\mathbf{y}_i = \mathbf{H}_i \mathbf{x}_i + \mathbf{n}_i \quad \mathbf{H}_i = \begin{bmatrix} \mathbf{h}_{i1} \\ \vdots \\ \mathbf{h}_{iN} \end{bmatrix} \quad i = 1 \dots N_c \quad (1)$$

where $N_c$ is the number of subcarriers, $\mathbf{x}_i$ and $\mathbf{y}_i$ are respectively the transmitted data vector and the received signal vector of size $N \times 1$, $\mathbf{H}_i$ the $N \times N$ MIMO channel matrix and $\mathbf{n}_i$ the vector containing colored noise. The transmitted vector is composed by $N$ vectors of size $N \times 1$ such that $\mathbf{x}_i = \sum_{j=1}^{N} \mathbf{q}_{ij}$.

$\mathbf{q}_{ij}$ is the vector of the $j^{\text{th}}$ user data where the $l^{\text{th}}$ element

of $\mathbf{q}_{ij}$ is the contribution of the $j^{\text{th}}$ user at the $i^{\text{th}}$ tone on the $l^{\text{th}}$ line. The second order moment of these data vectors are defined as $E[\mathbf{q}_{ij}\mathbf{q}_{ij}^H] = \mathbf{Q}_{ij}$. The covariance matrix of the transmitted data vector $E[\mathbf{x}_i\mathbf{x}_i^H] = \sum_{j=1}^{N} \mathbf{Q}_{ij}$ since the vectors $\mathbf{q}_{ij}$ are i.i.d. The dual MIMO Multiple Access Channel (MAC) for the dual uplink scenario with N users (see also Fig. 1) can be written as:

$$\mathbf{v}_i = \mathbf{H}_i^H \mathbf{u}_i + \mathbf{w}_i \quad \text{where} \quad \mathbf{H}_i^H = \begin{bmatrix} \mathbf{h}_{i1}^H & \dots & \mathbf{h}_{iN}^H \end{bmatrix} \quad (2)$$

where $\mathbf{u}_i = [u_{i1} \dots u_{iN}]^T$ is the transmitted vector on one tone $i$, $\mathbf{v}_i$ is the received signal vector of length $N$, and $\mathbf{w}_i$ is the vector containing colored noise. In this paper, we assume $E[\mathbf{n}_i\mathbf{n}_i^H] = \mathbf{I}$ (this is without loss of generality, as correlation in $\mathbf{n}_i$ cannot be exploited anyway), so that in the dual MAC channel $E[\mathbf{w}_i\mathbf{w}_i^H] = \mathbf{I}$ (in the MAC, a whitening operation can always be applied to the received vector such that the noise is white). The primal problem of finding optimal transmit vector covariance matrices in the BC under a total power constraint $P^{tot}$ is defined here as:

$$\max_{(\mathbf{Q}_{ij})_{i=1\dots N_c, j=1\dots N}} C^{BC}$$
$$\text{subject to} \sum_{j=1}^{N} \sum_{i=1}^{N_c} Trace(\mathbf{Q}_{ij}) \leq P^{tot} \quad (3)$$
$$\mathbf{Q}_{ij} \succeq 0, i = 1 \dots N_c, j = 1 \dots N$$

where $C^{BC}$ is the weighted rate sum function for MIMO BC employing DPC (as DPC achieve the MIMO BC capacity region [8]) for a given encoding order 1,...,N-1,N (i.e. user 1 is encoded first):

$$C^{BC} = \sum_{j=1}^{N} w_j \sum_{i=1}^{N_c} log_2 \left[ 1 + \mathbf{h}_{ij}\mathbf{Q}_{ij}\mathbf{h}_{ij}^H a_{ij}^{-1} \right] \quad (4)$$

where $a_{ij} = 1 + \mathbf{h}_{ij}(\sum_{k=j+1}^{N} \mathbf{Q}_{ik})\mathbf{h}_{ij}^H$. The $w_j$'s are the weights assigned to the different users. In the MAC, the optimal detection order is actually defined by the weights and the user with the largest weight is decoded last [10], [11]. Assuming a decreasing order of weights $w_1 > \cdots > w_K$, as the MAC-BC duality dictates a reverse of the decoding/encoding order, in the BC the user with the largest weight has indeed to be encoded first. Thus, the first term of the sum represents the rate of user 1, which is encoded under the crosstalk of the other users. The last term of the sum represents the rate of user N after having removed the crosstalk from the other users. The weighted rate sum function is neither convex nor concave [8], therefore finding the optimal transmit vector covariance matrices in the BC is a difficult task. Fortunately, the duality between the MAC and the BC states that it is possible to achieve the same set of rates in both domains under the same total power constraint. As the optimal power allocation in the MAC is tractable, one can calculate optimal transmit vector covariance matrices in the MAC and transform these into optimal transmit vector covariance matrices in the BC. The primal problem of finding power allocations in the MAC under a total power constraint $P^{tot}$ is:

$$\max_{(\mathbf{\Phi}_i)_{i=1\ldots N_c}} C^{MAC}$$
$$\text{subject to} \sum_{i=1}^{N_c} Trace(\mathbf{\Phi}_i) \leq P^{tot} \qquad (5)$$
$$\mathbf{\Phi}_i \succeq 0, i = 1 \ldots N_c$$

with $\mathbf{\Phi}_i = E[\mathbf{u}_i\mathbf{u}_i^H] = diag(\phi_{i1}, \ldots, \phi_{iN})$ the covariance matrix of transmitted symbols for tone $i$. The weighted rate sum function in the MAC for the decoding order N,N-1,...,1 (i.e. user 1 is decoded last) with Successive Interference Cancellation (SIC) is:

$$C^{MAC} = \sum_{j=1}^{N} w_j \sum_{i=1}^{N_c} log_2 \left[ det \left( \mathbf{I} + \mathbf{h}_{ij}^H \phi_{ij} \mathbf{h}_{ij} \mathbf{B}_{ij}^{-1} \right) \right] \qquad (6)$$

where $\mathbf{B}_{ij} = \mathbf{I} + \sum_{k=1}^{j-1} \mathbf{h}_{ik}^H \phi_{ik} \mathbf{h}_{ik}$. Problem (5)-(6) has been addressed in [10], [11]. The MAC-OSB algorithm has been derived based on a dual decomposition approach with a Lagrange formulation. First, the dual objective function of (5)-(6) is:

$$F^{MAC}(\lambda) = \max_{(\mathbf{\Phi}_i)_{i=1\ldots N_c}} \mathcal{L}^{MAC}(\lambda, (\mathbf{\Phi}_i)_{i=1\ldots N_c}) \qquad (7)$$

with $\lambda$ the Lagrange multiplier and

$$\mathcal{L}^{MAC}(\lambda, (\mathbf{\Phi}_i)_{i=1\ldots N_c}) = C^{MAC} + \lambda(P^{tot} - \sum_{i=1}^{N_c} Trace(\mathbf{\Phi}_i)) \qquad (8)$$

The dual optimization problem is:

$$\min_{\lambda} \quad F^{MAC}(\lambda)$$
$$\text{subject to} \quad \lambda \geq 0 \qquad (9)$$

By tuning the Lagrange multiplier, the total power constraint can be enforced. Because the dual objective function is concave with a convex constraint set, it has a unique minimum. As the duality gap is zero [8], [11], this minimum corresponds to the global optimum of the primal problem in (5)-(6). The search for the optimal $\lambda$ involves evaluations of the dual objective function (7), i.e. maximizations of the Lagrangian which, however, is decoupled over the tones for a given $\lambda$. In particular, the Lagrangian can be rewritten as:

$$\mathcal{L}^{MAC}(\lambda, (\mathbf{\Phi}_i)_{i=1\ldots N_c}) = \sum_{i=1}^{N_c} \left( \sum_{j=1}^{N} w_j log_2 \left[ \right. \right.$$
$$\left. \left. det \left( \mathbf{I} + \mathbf{h}_{ij}^H \phi_{ij} \mathbf{h}_{ij} \mathbf{B}_{ij}^{-1} \right) \right] - \lambda Trace(\mathbf{\Phi}_i) \right) + \lambda P^{tot} \qquad (10)$$

Therefore the maximization of the Lagrangian can be done by an exhaustive/iterative search on a per-tone basis [10] or by convex programming techniques such as interior point methods [11]. For further details, we refer to [10], [11]. Then, we can use the formulas from MAC-BC duality theory given in [8] to convert the optimal transmit vector covariance matrices found in the MAC domain into optimal transmit vector covariance matrices in the BC domain as follows:

---

for j=N to 1

- $a_{ij} = 1 + \mathbf{h}_{ij}(\sum_{k=j+1}^{N} \mathbf{Q}_{ik})\mathbf{h}_{ij}^H$  $(a_{iN} = 1)$

- $\mathbf{B}_{ij} = \mathbf{I} + \sum_{k=1}^{j-1} \mathbf{h}_{ik}^H \phi_{ik} \mathbf{h}_{ik}$  $(\mathbf{B}_{i1} = \mathbf{I})$

- $\mathbf{Q}_{ij} = \mathbf{B}_{ij}^{-1/2} \mathbf{F}_{ij} \mathbf{G}_{ij}^H a_{ij}^{1/2} \phi_{ij} a_{ij}^{1/2} \mathbf{G}_{ij} \mathbf{F}_{ij}^H \mathbf{B}_{ij}^{-1/2}$

end for

---

with $\mathbf{F}_{ij}$, $\mathbf{G}_{ij}$ the two unitary matrices coming from the SVD $\mathbf{B}_{ij}^{-1/2}\mathbf{h}_{ij}^H a_{ij}^{-1/2} = \mathbf{F}_{ij}\mathbf{L}_{ij}\mathbf{G}_{ij}^H$ and $\mathbf{L}_{ij}$ the diagonal matrix of singular values for user $j$ over tone $i$. Note that the above procedure requires that the same total transmit power is used for the power allocation in the MAC and for the transmit vector covariance matrices in the BC [8].

---

**Algorithm 1** Total power constraint

---

1 init $\lambda = 1$
2 init $step = 2$
3 init $b = 0$
4 init $(\mathbf{\Phi}_i)_{i=1\ldots N_c} = \mathbf{0}$
5 while $|\sum_{i=1}^{N_c} Trace(\mathbf{\Phi}_i) - P^{tot}| > \epsilon$
6     Exhaustive search $\max_{(\mathbf{\Phi}_i)_{i=1\ldots N_c}} \mathcal{L}^{MAC}(\lambda, (\mathbf{\Phi}_i)_{i=1\ldots N_c})$
7     if $\sum_{i=1}^{N_c} Trace(\mathbf{\Phi}_i) - P^{tot} < 0$
8         $b = b + 1$
9         $\lambda = \lambda/step$
10         $step = step - 1/2^b$
11     end if
12     $\lambda = \lambda * step$
13 end while
14 MAC-BC Duality
15 for j=N to 1 $\forall i = 1 \ldots N_c$
16     $a_{ij} = 1 + \mathbf{h}_{ij}(\sum_{k=j+1}^{N} \mathbf{Q}_{ik})\mathbf{h}_{ij}^H$
17     $\mathbf{B}_{ij} = \mathbf{I} + \sum_{k=1}^{j-1} \mathbf{h}_{ik}^H \phi_{ik} \mathbf{h}_{ik}$
18     $\mathbf{Q}_{ij} = \mathbf{B}_{ij}^{-1/2} \mathbf{F}_{ij} \mathbf{G}_{ij}^H a_{ij}^{1/2} \phi_{ij} a_{ij}^{1/2} \mathbf{G}_{ij} \mathbf{F}_{ij}^H \mathbf{B}_{ij}^{-1/2}$
19 end for

---

A complete algorithm description is given in **Algorithm 1**. We define $\epsilon$ as the tolerance between the actuel total power constraint $\sum_{i=1}^{N_c} Trace(\mathbf{\Phi}_i)$ and the target total power constraint $P^{tot}$. As the duality preserves the total power constraint, the transmit vector covariance matrices in the MAC (i.e. $\mathbf{\Phi}_i$) or the transmit vector covariance matrices in the BC (i.e. $\mathbf{Q}_{ij}$) can be used for convergence in the loop (line 5 in **Algorithm 1** can be replaced by $|\sum_{i=1}^{N_c}\sum_{j=1}^{N} Trace(\mathbf{Q}_{ij}) - P^{tot}| > \epsilon$). This algorithm leads to the same rates for the different users in the BC (4) and in the MAC (6) domain.

## III. BC POWER ALLOCATION UNDER PER-MODEM TOTAL POWER CONSTRAINTS

In the xDSL context, it is more relevant to consider a constraint on the transmit power of each modem separately instead of a constraint on the power for all modems together. Therefore, the goal of this section is to find optimal transmit vector covariance matrices in the BC under per-modem total power constraints, that is under a total power constraint for each modem over all tones. The primal problem is defined as:

$$\max_{(\mathbf{Q}_{ij})_{i=1\ldots N_c, j=1\ldots N}} C^{BC}$$
$$\text{subject to } \sum_{j=1}^{N} \sum_{i=1}^{N_c} \mathbf{Q}_{ij,ll} \leq P_l^{tot} \ \forall l \quad (11)$$
$$\mathbf{Q}_{ij} \succeq 0, i = 1 \ldots N_c, j = 1 \ldots N$$

with $P_l^{tot}$ the power budget for modem $l$, and $\mathbf{Q}_{ij,ll}$ the $l$th diagonal element of the transmit vector covariance matrix for user $j$ over tone $i$ coming from the $l$th diagonal element of the covariance matrix of the transmitted data vector $E[\mathbf{x}_i\mathbf{x}_i^H]_{ll} = \sum_{j=1}^{N} \mathbf{Q}_{ij,ll}$. Each diagonal element $\mathbf{Q}_{ij,ll}$ contributes to the per-modem total power constraint. $C^{BC}$ is the weighted rate sum function as defined by (4). We aim to follow a dual decomposition approach similar to the approach in section II, and again exploit MAC-BC duality theory. The primal problem of finding optimal power allocations in the MAC under a per-modem total power constraint $P_j^{tot}$ is:

$$\max_{(\mathbf{\Phi}_i)_{i=1\ldots N_c}} C^{MAC}$$
$$\text{subject to } \sum_{i=1}^{N_c} \phi_{ij} \leq P_j^{tot} \ \forall j \quad (12)$$
$$\mathbf{\Phi}_i \succeq 0, i = 1 \ldots N_c$$

with $\mathbf{\Phi}_i = E[\mathbf{u}_i\mathbf{u}_i^H] = diag(\phi_{i1}, \ldots, \phi_{iN})$ the transmit covariance matrix for tone i and $C^{MAC}$ the weighted rate sum defined by (6). However, the MAC optimal power allocation computed from (12) cannot be converted directly into BC optimal transmit vector covariance matrices for (11) because the MAC-BC duality does not preserve per-modem total power constraints. To bypass this problem we apply a transformation to the dual objective function of (11) leading to an equivalent objective function with a total power constraint, and then we exploit MAC-BC duality. This transformation consists of a rescaling of the channel matrices by a virtual precoding matrix operation. The BC dual objective function corresponding to (11) is:

$$F^{BC}(\mathbf{\Lambda}) = \max_{(\mathbf{Q}_{ij})_{i=1\ldots N_c, j=1\ldots N}} \mathcal{L}^{BC}(\mathbf{\Lambda}, (\mathbf{Q}_{ij})_{i=1\ldots N_c, j=1\ldots N}) \quad (13)$$

with

$$\mathcal{L}^{BC}(\mathbf{\Lambda}, (\mathbf{Q}_{ij})_{i=1\ldots N_c, j=1\ldots N}) = \sum_{i=1}^{N_c} \left( \sum_{j=1}^{N} w_j log_2 \Big[ 1 + \mathbf{h}_{ij}\mathbf{Q}_{ij}\mathbf{h}_{ij}^H a_{ij}^{-1} \Big] - \sum_{j=1}^{N} Trace(\mathbf{\Lambda}\mathbf{Q}_{ij}) \right) + Trace\left(\mathbf{\Lambda} diag(P_1^{tot}, \ldots, P_L^{tot})\right) \quad (14)$$

with $L$ the number of modems used in the downstream ($L$ can be larger or equal than the number of active users $N$)[1], and $\mathbf{\Lambda}$ a diagonal matrix of Lagrange multipliers $diag(\lambda_1, \ldots, \lambda_L)$. The dual optimization problem is:

$$\text{minimize} \quad F^{BC}(\mathbf{\Lambda})$$
$$\text{subject to} \quad \lambda_l \geq 0 \quad \forall l \quad (15)$$

First we will solve the dual problem (15). Then we will show zero duality gap between the primal and the dual problems in (11) and (15) respectively. Rescaling the channel matrices by the inverse square root of the Lagrange multiplier matrix leads to:

$$\mathbf{y}_i = \overbrace{\mathbf{H}_i\mathbf{\Lambda}^{-1/2}}^{\mathbf{H}'_i}\overbrace{\mathbf{\Lambda}^{1/2}\mathbf{x}_i}^{\mathbf{x}'_i} + \mathbf{n}_i. \quad (16)$$

For this equivalent channel and a given $\mathbf{\Lambda}$, the dual objective function in the BC becomes:

$$F^{BC}(\mathbf{\Lambda}) = \max_{(\mathbf{Q}'_{ij})_{i=1\ldots N_c, j=1\ldots N}} \mathcal{L}^{BC}(\mathbf{\Lambda}, (\mathbf{Q}'_{ij})_{i=1\ldots N_c, j=1\ldots N}) \quad (17)$$

with

$$\mathcal{L}^{BC}(\mathbf{\Lambda}, (\mathbf{Q}'_{ij})_{i=1\ldots N_c, j=1\ldots N}) = \sum_{i=1}^{N_c} \left( \sum_{j=1}^{N} w_j log_2 \Big[ \Big(1 + \mathbf{h}'_{ij}\mathbf{Q}'_{ij}\mathbf{h}'^H_{ij} a'^{-1}_{ij}\Big) \Big] - \sum_{j=1}^{N} Trace(\mathbf{Q}'_{ij}) \right) + Trace\left(diag(P'^{tot}_1, \ldots, P'^{tot}_L)\right) \quad (18)$$

where $a'_{ij} = 1 + \mathbf{h}'_{ij}(\sum_{k=j+1}^{N} \mathbf{Q}'_{ik})\mathbf{h}'^H_{ij}$ with the rescaled channel vectors $\mathbf{h}'_{ij} = \mathbf{h}_{ij}\mathbf{\Lambda}^{-1/2}$ and the re-defined transmit covariance matrices $\mathbf{Q}'_{ij} = \mathbf{\Lambda}^{1/2}\mathbf{Q}_{ij}\mathbf{\Lambda}^{1/2}$. The target per-modem total power constraints are also re-defined as $diag(P'^{tot}_1, \ldots, P'^{tot}_L) = \mathbf{\Lambda}^{1/2}diag(P_1^{tot}, \ldots, P_L^{tot})\mathbf{\Lambda}^{1/2}$. One can see that (18) corresponds to (10) with $\lambda = 1$, and so that the precoder matrix $\mathbf{\Lambda}^{-1/2}$ transforms the per-modem total power constraints into a virtual total power constraint by hiding the Lagrange multipliers into the equivalent channels $\mathbf{h}'_{ij}$ and the new covariance matrices $\mathbf{Q}'_{ij}$. We can now invoke MAC-BC duality theory to transform the dual optimization problem in the BC into a dual optimization problem in the MAC for a given $\mathbf{\Lambda}$ under the same virtual total power constraint. The MAC-BC duality theory states that the sum rate capacity of the MIMO BC equals the sum rate capacity of the MIMO MAC with preservation of the total power constraint. Therefore $F^{BC}(\mathbf{\Lambda})$ in (13) is identical to $F^{MAC}(\mathbf{\Lambda})$ in (22) resulting from the following equality concerning the sum rate capacity:

$$
\overbrace{\sum_{j=1}^{N} w_j \sum_{i=1}^{N_c} log_2\Big[\Big(1 + \mathbf{h'}_{ij}\mathbf{Q'}_{ij}\mathbf{h'}_{ij}^{H}{a'}_{ij}^{-1}\Big)\Big]}^{BC} =
$$
$$
\underbrace{\sum_{j=1}^{N} w_j \sum_{i=1}^{N_c} log_2\Big[det\Big(\mathbf{I} + \mathbf{h'}_{ij}^{H}\phi'_{ij}\mathbf{h'}_{ij}\mathbf{B'}_{ij}^{-1}\Big)\Big]}_{MAC} \quad (19)
$$

where $\mathbf{B'}_{ij} = \mathbf{I} + \sum_{k=1}^{j-1} \mathbf{h'}_{ik}^{H}\phi'_{ik}\mathbf{h'}_{ik}$. The preservation of the total power constraint leads to the following equality :

$$
\sum_{i=1}^{N_c}\sum_{j=1}^{N} Trace(\mathbf{Q'}_{ij}) = \sum_{i=1}^{N_c} Trace(\mathbf{\Phi'}_i) \quad (20)
$$

---

[1]For example other lines that are not dedicated to specific users (or that are not active) are used to boost the rate of the active users through crosstalk coupling.

Based on the two equalities (19) and (20), the BC dual optimization problem in (15) can be rewritten as:

$$
\begin{aligned}
\text{minimize} \quad & F^{MAC}(\mathbf{\Lambda}) \\
\text{subject to} \quad & \lambda_l \geq 0 \quad \forall l
\end{aligned} \quad (21)
$$

with

$$
F^{MAC}(\mathbf{\Lambda}) = \max_{(\mathbf{\Phi'}_i)_{i=1\dots N_c}} \mathcal{L}^{MAC}(\mathbf{\Lambda}, (\mathbf{\Phi'}_i)_{i=1\dots N_c}) \quad (22)
$$

and where $\mathcal{L}^{MAC}$ is the same as $\mathcal{L}^{BC}$ in (18) except that we take into account (19) and (20) such that

$$
\begin{aligned}
\mathcal{L}^{MAC}(\mathbf{\Lambda}, (\mathbf{\Phi'}_i)_{i=1\dots N_c}) = \sum_{i=1}^{N_c}\Bigg(\sum_{j=1}^{N} w_j log_2\Big[ \\
det\Big(\mathbf{I} + \mathbf{h'}_{ij}^{H}\phi'_{ij}\mathbf{h'}_{ij}\mathbf{B'}_{ij}^{-1}\Big)\Big] - Trace(\mathbf{\Phi'}_i))\Bigg) \\
+ Trace\Big(diag(P'^{tot}_1,\dots,P'^{tot}_L)\Big)
\end{aligned} \quad (23)
$$

Therefore, for a given $\mathbf{\Lambda}$, we can then compute the optimal power allocation in the MAC by (22) and use the duality formulas of [8] to obtain the optimal transmit vector covariance matrices in (13). The Lagrange multipliers are then adjusted so that the per-modem total power constraints are enforced using (15).

Up to here we found the optimal solution of the dual problem (15) by means of (21). In the following we will demonstrate that the duality gap between (11) and (15) is zero. Towards this end, we introduce the following theorem:

*Theorem:* The primal optimization problem in the BC under per-modem total power constraints (11) is identical to the dual optimization problem in the BC under per-modem total power constraints (15). Hereby duality gap between (11) and (15) is zero.

*Proof*: It is known from [9] that the capacity region of a BC under per-antenna power constraints for a multi-tone transmission is identical to the capacity region of a MAC under a total power constraint with an uncertain noise diagonal

matrix. We can write the equivalent primal problem in the BC as:

$$
\begin{aligned}
\min_{\mathbf{X}} \max_{(\mathbf{\Phi}_i)_{i=1\dots N_c}} \sum_{j=1}^{N} w_j \sum_{i=1}^{N_c} log_2\Big[det\Big(\mathbf{I} + \mathbf{h}_{ij}^{H}\phi_{ij}\mathbf{h}_{ij}\mathbf{Y}_{\mathbf{X},ij}^{-1}\Big)\Big] \\
\text{subject to} \sum_{i=1}^{N_c} Trace(\mathbf{\Phi}_i) \leq P^{tot} \\
\mathbf{\Phi}_i \succeq 0, i = 1\dots N_c
\end{aligned} \quad (24)
$$

where $\mathbf{Y}_{\mathbf{X},ij} = \mathbf{X} + \sum_{k=1}^{j-1} \mathbf{h}_{ik}^{H}\phi_{ik}\mathbf{h}_{ik}$, $\mathbf{X}$ is an uncertain noise diagonal matrix and $P^{tot}$ is the sum of the per-modem total power constraints in the BC. As the primal and the dual optimization problems of MAC under a total power constraint lead to the same capacity region (the duality gap is zero since the problem has a concave objective function under a convex constraint set), for a given $\mathbf{X}$ the dual formulation under a total power constraint:

$$
\begin{aligned}
\min_{\mathbf{X},\lambda} \max_{(\mathbf{\Phi}_i)_{i=1\dots N_c}} \sum_{i=1}^{N_c}\Bigg(\sum_{j=1}^{N} w_j log_2\Big[det\Big(\mathbf{I} + \mathbf{h}_{ij}^{H}\phi_{ij}\mathbf{h}_{ij}\mathbf{Y}_{\mathbf{X},ij}^{-1}\Big)\Big] \\
- \lambda \sum_{i=1}^{N_c} Trace(\mathbf{\Phi}_i)\Bigg) + \lambda P^{tot}
\end{aligned} \quad (25)
$$

Now, taking into account the previous definitions of the new covariance matrices $\lambda\mathbf{\Phi}_i = \mathbf{\Phi}_i^*$ and new total power constraint $\lambda P^{tot} = P^{*tot}$, we get:

$$
\begin{aligned}
\min_{\mathbf{X},\lambda} \max_{(\mathbf{\Phi}_i^*)_{i=1\dots N_c}} \sum_{i=1}^{N_c}\Bigg(\sum_{j=1}^{N} w_j log_2\Big[det\Big(\mathbf{I} + \frac{1}{\lambda}\mathbf{h}_{ij}^{H}\phi_{ij}^{*}\mathbf{h}_{ij}\mathbf{Y}_{\mathbf{X},ij}^{*-1}\Big)\Big] \\
- \sum_{i=1}^{N_c} Trace(\mathbf{\Phi}_i^*)\Bigg) + P^{*tot}
\end{aligned} \quad (26)
$$

with $\mathbf{Y}_{\mathbf{X},ij}^* = \mathbf{X} + \frac{1}{\lambda}\sum_{k=1}^{j-1} \mathbf{h}_{ik}^{H}\phi_{ik}^{*}\mathbf{h}_{ik}$. The interference can be rewritten as:

$$
\mathbf{Y}_{\mathbf{X},ij}^* = \mathbf{X}^{1/2}(\mathbf{I} + \frac{1}{\lambda}\sum_{k=1}^{j-1}\mathbf{X}^{-1/2}\mathbf{h}_{ik}^{H}\phi_{ik}^{*}\mathbf{h}_{ik}\mathbf{X}^{-1/2})\mathbf{X}^{1/2} \quad (27)
$$

Using the following equality $det(\mathbf{I} + \mathbf{AB}) = det(\mathbf{I} + \mathbf{BA})$, this leads to the following optimization problem:

$$
\begin{aligned}
\min_{\mathbf{X},\lambda} \max_{(\mathbf{\Phi}_i^*)_{i=1\dots N_c}} \sum_{i=1}^{N_c}\Bigg(\sum_{j=1}^{N} w_j log_2\Big[ \\
det\Big(\mathbf{I} + \frac{1}{\lambda}\mathbf{X}^{-1/2}\mathbf{h}_{ij}^{H}\phi_{ij}^{*}\mathbf{h}_{ij}\mathbf{X}^{-1/2}\mathbf{B}_{\mathbf{X},ij}^{*-1}\Big)\Big] \\
- \sum_{i=1}^{N_c} Trace(\mathbf{\Phi}_i^*)\Bigg) + P^{*tot}
\end{aligned} \quad (28)
$$

with $\mathbf{B}_{\mathbf{X},ij}^* = \mathbf{I} + \frac{1}{\lambda}\sum_{k=1}^{j-1} \mathbf{X}^{-1/2}\mathbf{h}_{ik}^{H}\phi_{ik}^{*}\mathbf{h}_{ik}\mathbf{X}^{-1/2}$. When the uncertain noise is selected such that $\frac{1}{\sqrt{\lambda}}\mathbf{X}^{-1/2} = \mathbf{\Lambda}^{-1/2}$ and knowing that $\mathbf{h'}_{ij} = \mathbf{h}_{ij}\mathbf{\Lambda}^{-1/2}$, one can easily show that $\mathbf{B}_{\mathbf{X},ij}^* = \mathbf{B'}_{ij}$ with $\mathbf{\Phi}^* = \mathbf{\Phi'}$. Hence (28) will be exactly the same optimization problem as (21). Knowing that the (28), (21) and (15) optimization problems are the same, the duality gap between (24) and (28) is zero, and (24) has the same capacity region as (11), we can conclude that the duality gap between (11) and (15) is zero. This completes the proof.

A complete algorithm description is given as **Algorithm 2**. We define $\epsilon_l$ as the tolerance between the actuel per-modem total power constraint $\sum_{i=1}^{N_c} [\mathbf{\Lambda}^{-1/2}(\sum_{j=1}^{N}\mathbf{Q}'_{ij})\mathbf{\Lambda}^{-1/2}]_{ll}$ and the target per-modem total power constraint $P_l^{tot}$ for the $l^{th}$ modem. This algorithm will be referred to as BC-OSB (BC-Optimal Spectrum Balancing):

---

**Algorithm 2** BC-OSB under per-modem total power constraints

---

1  for $l = 1 \ldots L$
2      init $\lambda_l = 1$
3      init $step_l = 2$
4      init $b_l = 0$
5      init $(\mathbf{Q}'_{ij})_{i=1\ldots N_c, j=1\ldots N} = \mathbf{0}$
6  end for
7  while $\exists l$ s.t. $|\sum_{i=1}^{N_c}[\mathbf{\Lambda}^{-1/2}(\sum_{j=1}^{N}\mathbf{Q}'_{ij})\mathbf{\Lambda}^{-1/2}]_{ll} - P_l^{tot}| > \epsilon_l$
8      Exhaustive search $\max_{(\mathbf{\Phi}'_i)_{i=1\ldots N_c}} \mathcal{L}^{MAC}(\mathbf{\Lambda}, (\mathbf{\Phi}'_i)_{i=1\ldots N_c})$
9      MAC-BC Duality
10     for j=N to 1 $\forall i = 1 \ldots N_c$
11         $a'_{ij} = 1 + \mathbf{h}'_{ij}(\sum_{k=j+1}^{N}\mathbf{Q}'_{ik})\mathbf{h}'^H_{ij}$
12         $\mathbf{B}'_{ij} = \mathbf{I} + \sum_{k=1}^{j-1}\mathbf{h}'^H_{ik}\phi'_{ik}\mathbf{h}'_{ik}$
13         $\mathbf{Q}'_{ij} = \mathbf{B}'^{-1/2}_{ij}\mathbf{F}'_{ij}\mathbf{G}'^H_{ij}a'^{1/2}_{ij}\phi'_{ij}a'^{1/2}_{ij}\mathbf{G}'_{ij}\mathbf{F}'^H_{ij}\mathbf{B}'^{-1/2}_{ij}$
14     end for
15     for $l = 1 \ldots L$
16         if $\sum_{i=1}^{N_c}[\mathbf{\Lambda}^{-1/2}(\sum_{j=1}^{N}\mathbf{Q}'_{ij})\mathbf{\Lambda}^{-1/2}]_{ll} - P_l^{tot} < 0$
17             $b_l = b_l + 1$
18             $\lambda_l = \lambda_l/step_l$
19             $step_l = step_l - 1/2^{b_l}$
20         end if
21         $\lambda_l = \lambda_l * step_l$
22     end for
23 end while

---

**TABLE I**

**RATES UNDER TOTAL POWER CONSTRAINT IN THE BC.**

| R1/R2 (Mbps) | $w_1$=0.0 | $w_1$=0.1 | $w_1$=0.2 |
|---|---|---|---|
| User 1 first | 0/95.93 | 120.9/94.99 | 127.3/93.92 |
| User 2 first | 0/95.93 | 120.9/94.99 | 127.3/93.93 |

| $w_1$=0.3 | $w_1$=0.4 | $w_1$=0.5 | $w_1$=0.6 |
|---|---|---|---|
| 131.6/92.53 | 133.7/91.41 | 136.2/89.37 | 137.9/87.18 |
| 131.6/92.53 | 133.7/91.41 | 136.2/89.37 | 137.9/87.18 |

| $w_1$=0.7 | $w_1$=0.8 | $w_1$=0.9 | $w_1$=1.0 |
|---|---|---|---|
| 139.1/85.05 | 140.4/80.80 | 141.5/74.44 | 142.4/0 |
| 139.1/85.05 | 140.4/80.80 | 141.5/74.44 | 142.4/0 |

**TABLE II**

**RATES UNDER PER-MODEM TOTAL POWER CONSTRAINT IN THE BC.**

| R1/R2(Mbps) | $w_1$=0.0 | $w_1$=0.1 | $w_1$=0.2 |
|---|---|---|---|
| User 1 first | 0/89.96 | 136.0/89.53 | 136.0/89.53 |
| User 2 first | 0/89.96 | 136.0/89.53 | 136.0/89.53 |

| $w_1$=0.3 | $w_1$=0.4 | $w_1$=0.5 | $w_1$=0.6 |
|---|---|---|---|
| 136.0/89.53 | 136.0/89.53 | 136.0/89.53 | 136.0/89.53 |
| 136.0/89.53 | 136.0/89.53 | 136.0/89.53 | 136.0/89.53 |

| $w_1$=0.7 | $w_1$=0.8 | $w_1$=0.9 | $w_1$=1.0 |
|---|---|---|---|
| 136.0/89.53 | 136.0/89.53 | 136.0/89.53 | 136.2/0 |
| 136.0/89.53 | 136.0/89.53 | 136.0/89.53 | 136.2/0 |

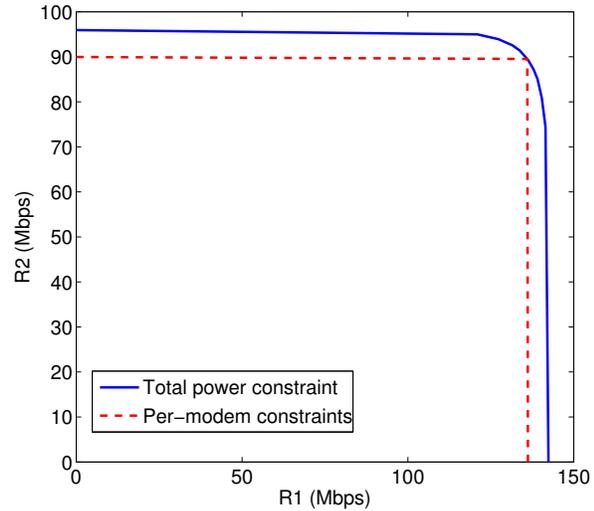

Fig. 2. Rate region of BC-OSB in a VDSL2 system under a total power constraint and per-modem total power constraints.

## IV. RESULTS

In this section, we provide simulation results for two different VDSL2 downlink scenarios. We use measured channels from a France Telecom binder. In the first scenario, 2 users respectively at 400 and 800 meters from the Central Office (CO) or the Remote Terminal (RT) are served with Differential-Mode (DM) lines ($2\times2$ channel matrix). In the second scenario, 2 users both at 400 meters from the CO/RT are served with Differential-Mode (DM) lines and their Phantom-mode (PM) line (differential between the 2 common modes) giving a $2\times3$ channel matrix. In this scenario, external noise is coming from 2 other DM lines. The spectral masks for VDSL2 Fiber-To-The-exchange (FTTex) are applied [12], with SNR gap $\Gamma$=0 dB (since MAC-BC duality does not hold if $\Gamma >0$ dB)[2], an AWGN of -140 dBm/Hz and a maximum transmit power $P_j^{tot}$=14.5 dBm per line. The frequency range is from 0 to

[2]The more general case with $\Gamma >0$ dB will be addressed in a future report. The transmit covariance matrices $(\mathbf{Q}_{ij})_{i=1\ldots N_c, j=1\ldots N}$ are optimal transmit covariance matrices optimal for the DPC capacity formula. Note that from the implementation point of view, once the optimal covariance matrices $(\mathbf{Q}_{ij})_{i=1\ldots N_c, j=1\ldots N}$ are determined, the transmitted data symbols $\mathbf{x}_i$ can be constructed as follows:

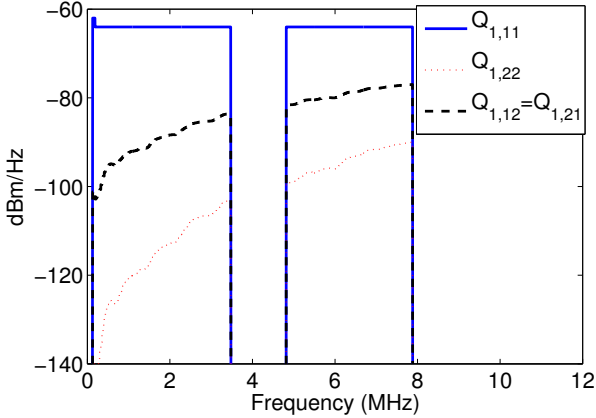

Fig. 3. Optimal BC covariance matrix for user 1 in a VDSL2 scenario under per-modem total power constraints.

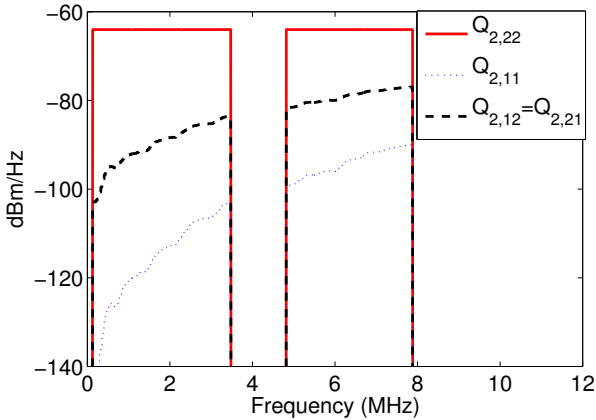

Fig. 4. Optimal BC covariance matrix for user 2 in a VDSL2 scenario under per-modem total power constraints

12 MHz with 4.3125 kHz spacing between subcarriers and 4 kHz symbol rate. The FDD band plan of VDSL2 up to 12 MHz provides 2 frequency bands in the downlink scenario, namely 138kHz-3.75MHz and 5.2MHz-8.5MHz. In the second scenario, we transmit in the 0-30 MHz range with the FDD band plan of VDSL2 up to 12 MHz and transmitting in the all 12-30 MHz bandwidth with similar per-modem total power constraints.

For TABLE I a total power constraint $P^{tot} = 29$ dBm for the 2 modems is chosen. The table shows the rates obtained in the 2×2 BC with different sets of weights $(w_1, w_2) = (w_1, 1 - w_1)$. The two lines refer to two different encoding orders in

1) The $N \times 1$ vector of the M-QAM data symbols $\mathbf{s}_{ij}$ is precoded using the $N \times N$ $\mathbf{L}_{ij}$ matrix, i.e. $\mathbf{q}_{ij} = \mathbf{L}_{ij}\mathbf{s}_{ij}$, such that $E[\mathbf{q}_{ij}\mathbf{q}_{ij}^H] = \mathbf{Q}_{ij} = \mathbf{L}_{ij}\mathbf{L}_{ij}^H$ from Cholesky decomposition (as $\mathbf{Q}_{ij}$ is a positive semi-definite matrix).

2) Then $\mathbf{x}_i = \sum_{j=1}^{N} \mathbf{q}_{ij}$ will be sent on the $N$ lines (the $l^{th}$ element of $\mathbf{x}_i$ will be sent on the $l^{th}$ line).

the BC. The rates R1 and R2 of the two users are provided, with the first line corresponding to user 1 encoded first and the second line corresponding to user 2 encoded first. We can see that the larger the weight allocated to one user, the larger the rate allocated to this user. However, one can see that the difference between the possible encoding orders is negligible (approximately $10^{-3}$ Mbps) due to the diagonal dominance of the channel matrix.

For TABLE 2 per-modem total power constraints $P_j^{tot} = 14.5$ dBm are chosen. The table again shows the rates obtained in the 2×2 BC. As in the total power constraint case, the results show almost equal rates for any encoding order, due to the diagonal dominance of the channel matrix. Moreover, even varying the weights do not affect the different rates of the users (except for the extreme cases $w_1$=0 and $w_1$=1). In fact, the resulting transmit covariance matrices do not depend on the weights owing to the precoding matrix operation which rescales the channel vectors in the same way for these different weights.

In Fig. 2, we plotted the rates of the two first tables. The Diagonalizing Precoder (DP) of [13] achieves 135.8 Mbps for the first user and 89.30 Mbps for the second user. We can conclude that in this scenario, owing to the diagonal dominance of the channel matrix, the DP achieves most of the capacity. Fig. 3 and Fig. 4 show the two BC optimal transmit vector covariance matrices $\mathbf{Q}_j$ as the covariance matrix where the tone dependency was removed. Its $(m, n)^{th}$ entry is denoted $\mathbf{Q}_{j,mn}$ over all frequency tones $i$ under per-modem total power constraints with $m$ and $n$ referring to the $m^{th}$ row and the $n^{th}$ column of the matrix $\mathbf{Q}_{ij,mn}$. We can see that the optimal transmit covariance matrices have an almost flat power allocation in the direct channels and a much smaller power allocation in the crosstalk channels, where the profile follows the shape of the FEXT.

| Sum rate (Mbps) | SVD TPC [14] | SVD PMTPC [14] |
|---|---|---|
| 2x2 | 461.46 | 461.46 |
| 2x3 | 488.64 | 469.97 |

| DP PMTPC [13] | BC-OSB PMTPC |
|---|---|
| 448.47 | 449.01 |
| 449.01 | 458.94 |

TABLE III
COMPARISON BETWEEN DIFFERENT SCHEMES

The second set of simulation results involves a VDSL2 scenario with a 0 to 30 MHz bandwidth and exploiting 2 DM lines and their PM. In this case, the 2 DM lines are used for downlink transmission with external noise coming from 2 VDSL2 DM lines of similar length, whose PSD's are set at -60 dBm/Hz. The PM is also used at the transmit side for downlink transmission, providing an overall channel matrix of size 2×3. By duality, this corresponds to a 3×2 MAC case where 2 users are transmitting in an uplink scenario and the 2 DM lines and PM are used at the receive side. We use per-modem total power constraints with 14.5 dBm per modem. We transmit in the 0-30 MHz range with the FDD band plan of VDSL2 up to 12 MHz and transmitting in the all 12-30 MHz bandwidth with similar per-modem total power constraints.

TABLE 3 shows a comparison between existing algorithms [13], [14] for the $2\times2$ case without exploiting the PM and the $2\times3$ case exploiting the PM. The weights are $w1 = w2 = 0.5$. The algorithms SVD under a Total Power Constraint (TPC) and SVD under Per-Modem Total Power Constraints (PMTPC) provide the optimal rate sum with two-sided coordination [14]. Owing to the external noise, there is a rate loss between the SVD schemes (with two-sided coordination) and the DP PMTPC or the BC-OSB PMTPC (with only transmit side coordination). The DP and the BC-OSB algorithms lead to the same rates in the $2\times2$ case because none of these algorithms can properly mitigate the external noise (whitening operation not possible at the receive side). However, in the $2\times3$ case, the BC-OSB shows a increased rate compared to the DP. This is also due to the linear structure of the DP which make it difficult to meet the per-modem total power constraints for non-square matrices.

Simulation results were performed on xDSL systems, but they can also apply to wireless systems. The two main differences between xDSL systems and wireless systems is that the former exhibit diagonal dominance of the channel matrix while the latter do not. Therefore, in xDSL systems, while linear precoders can achieve most of the capacity, wireless systems require more advanced processing algorithms (such as the BC-OSB algorithm presented in this paper) in order to achieve the maximum sum rate capacity. An extension of this work is to investigate the BC-OSB algorithm in wireless systems.

## V. Conclusion

In this paper we have investigated the problem of optimal power allocation in the MIMO BC in the context of downstream xDSL. We have first described an algorithm for power allocation under a total power constraint, i.e a single total power constraint for all modems over all tones. Then, a new algorithm called BC-OSB algorithm has been devised for a more realistic power allocation under per-modem total power constraints, i.e a total power constraint for each modem over all tones, where the derivation is based on a dual problem formulation and an adequate transformation by a precoding matrix. Simulation results have been provided for different scenarios, namely a VDSL2 scenario with DM transmission and a VDSL2 scenario with DM and PM transmission.

## References

[1] V. L. Nir, M. Moonen, J. Verlinden, and M. Guenach, "Broadcast channel optimum spectrum balancing (bc-osb) with per-modem total power constraints for downstream dsl," in *15th European Signal Processing Conference (EUSIPCO'2007), September 4-8, 2007, Poznan, Poland*, September 2007.

[2] G. J. Foschini and M. J. Gans, "On limits of wireless communications in a fading environment when using multiple antennas," *Wireless Personal Communications*, vol. 6, no. 3, pp. 311–335, 1998.

[3] E. Telatar, "Capacity of multi-antenna gaussian channels," *European Trans. on Telecomm. ETT*, vol. 10, no. 6, pp. 585–596, Nov. 1999.

[4] W. Yu, W. Rhee, S. Boyd, and J. Cioffi, "Iterative water-filling for gaussian vector multiple-access channels," *IEEE transaction on Information Theory*, vol. 50, no. 1, pp. 145–152, Jan. 2004.

[5] T. Cover and J. Thomas, *Elements of Information Theory*. New York: Wiley, 1991.

[6] G. Caire and S. shamai, "On the achievable throughput of a multi-antenna gaussian broadcast channel," *IEEE Transactions on Vehicular Technology*, vol. 49, pp. 1691–1706, July 2003.

[7] H. Weingarten, Y. Steinberg, and S. Shamai, "The capacity region of the gaussian mimo broadcast channel," in *Conference on Information Sciences and Systems*, 2004, pp. 7–12.

[8] S. Vishwanath, N. Jindal, and A. Goldsmith, "Duality, achievable rates, and sum-rate capacity of gaussian mimo broadcast channels," *IEEE Transactions on Information Theory*, vol. 49, no. 10, pp. 2658–2668, October 2003.

[9] W. Yu and T. Lan, "Transmitter optimization for the multi-antenna downlink with per-antenna power constraints," *IEEE Transactions on Signal Processing*, vol. 55, no. 6, pp. 2646–2660, June 2007.

[10] P. Tsiaflakis, J. Vangorp, J. Verlinden, and M. Moonen, "Multiple access channel optimal spectrum balancing for upstream dsl transmission," *IEEE Communication Letters*, vol. 11, no. 4, pp. 398–400, Apr. 2007.

[11] H. Boche and M. Wiczanowski, "Optimization-theoretic analysis of stability-optimal transmission policy for multiple-antenna multiple-access channel," *IEEE Transactions on Signal Processing*, vol. 55, no. 6, pp. 2688–2702, June 2007.

[12] G.993.2, "Very high speed digital subscriber line transceivers 2 (VDSL2)," *ITU-T Recommendation*, Feb. 2006.

[13] R. Cendrillon, G. Ginis, E. V. den Bogaert, and M. Moonen, "A near-optimal linear crosstalk precoder for downstream vdsl," *IEEE Transactions on Communications*, vol. 55, no. 5, p. 860863, May 2007.

[14] V. L. Nir, M. Moonen, and J. Verlinden, "Optimal power allocation under per-modem total power and spectral mask constraints in xdsl vector channels with alien crosstalk," in *IEEE International Conference on Acoustics, Speech, and Signal Processing, ICASSP'07, Honolulu, USA*, April 2007.